\documentclass[]{spie}  

\usepackage{amsmath,amsfonts,amssymb}
\usepackage{graphicx}
\usepackage[colorlinks=true, allcolors=blue]{hyperref}

\title{The Space Coronagraph Optical Bench (SCoOB): 7. design, fabrication, and first light for a self-coherent camera}

\author[a]{Kevin Derby}
\author[a]{Kian Milani}
\author[a]{G.C. Hathaway}
\author[a]{Joshua Liberman}
\author[b]{Kyle Van Gorkom}
\author[b]{Ramya Anche}
\author[b]{Adam Schilperoort}
\author[b]{Corey Fucetola}
\author[a]{Brandon Chalifoux}
\author[c]{Kuravi Hewawasam}
\author[c]{Christopher Mendillo}
\author[d]{Sebastiaan Y. Haffert}
\author[b]{Ewan S. Douglas}

\affil[a]{James C. Wyant College of Optical Sciences, University of Arizona, 1630 E University Blvd, Tucson, AZ 85721, USA}
\affil[b]{Steward Observatory, University of Arizona, 933 N Cherry Ave, Tucson, AZ 85721, USA}
\affil[c]{Lowell Center for Space Science \& Technology, University of Massachusetts Lowell, 220 Pawtucket St, Lowell, MA 01854, USA}
\affil[d]{Universiteit Leiden, Rapenburg 70, 2311 EZ Leiden, Netherlands}

\authorinfo{Further author information: (Send correspondence to E.S.D.)\\E.S.D.: E-mail: douglase@arizona.edu}

\begin{document} 
\maketitle

\begin{abstract}
The 2020 Decadal Survey on Astronomy and Astrophysics tasked future space observatories with the goal of detecting and characterizing a large sample of Earth-like exoplanets. To achieve this, these observatories will require coronagraphs and wavefront control algorithms in order to achieve 10\textsuperscript{-10} or better starlight suppression. The Space Coronagraph Optical Bench (SCoOB) is a vacuum compatible testbed at the University of Arizona which aims to advance and mature starlight suppression technologies in a space-like environment. In its current configuration, SCoOB is a charge-6 vector vortex coronagraph outfitted with a Kilo-C microelectromechanical systems deformable mirror capable of achieving sub-10\textsuperscript{-8} dark hole contrast at visible wavelengths using implicit electric field conjugation (iEFC). 

In this work, we demonstrate the use of a self-coherent camera (SCC) for dark hole digging and maintenance on SCoOB. The SCC introduces a small off-axis pinhole in the Lyot plane which allows some starlight to reach the focal plane and interfere with residual speckles. This enables high-order focal-plane wavefront sensing which can be combined with active wavefront control to null the speckles in a specified region of high contrast known as the dark hole. We discuss considerations for implementation, potential limitations, and provide a performance comparison with iEFC. We also discuss the design optimization and fabrication process for our SCC Lyot stops. 
\end{abstract}


\section{INTRODUCTION}
\label{sec:intro}

Future observatories with the goal of detecting and characterizing a large sample of Earth-like exoplanets will require advanced starlight suppression techniques to reach the 10\textsuperscript{-10} contrast necessary to image these targets in reflected light\cite{deeg_direct_2018, guyon_extreme_nodate}. Coronagraphs provide an attractive solution for starlight suppression by rejecting on-axis starlight using a combination of pupil and focal plane masks. However, residual starlight caused by imperfections in the optical system create speckles at the science camera, which can be orders of magnitude brighter than any exoplanet. To overcome this, we measure the speckles and command a deformable mirror (DM) to null them in a specified region of high contrast known as the dark hole (DH) using high-order wavefront sensing and control (HOWFSC)\cite{giveon_electric_2007}.

\begin{figure} [h!]
\centering\includegraphics[width=12cm]{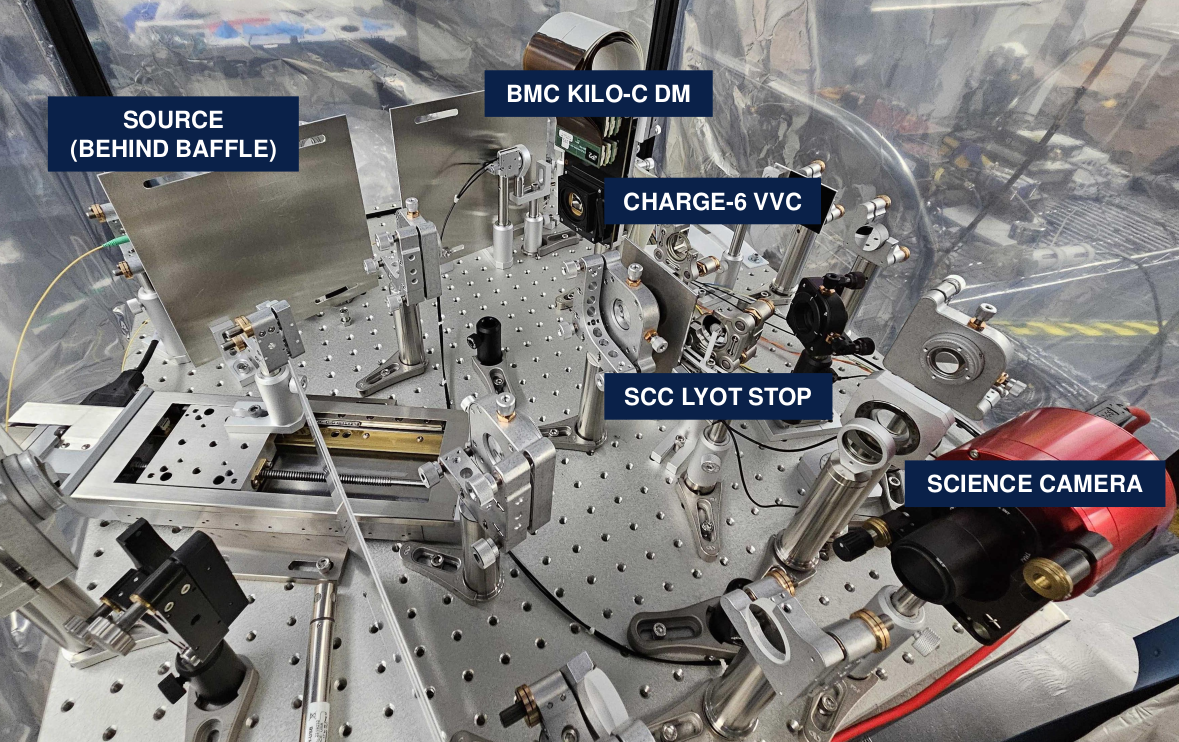}
\caption{The Space Coronagraph Optical Bench at the University of Arizona. The 3D-printed flag used for modulating the SCC pinhole can be seen. This work was performed with the testbed in-air in the configuration shown here.} 
\label{fig:scoob} 
\end{figure} 

The Space Coronagraph Optical Bench (SCoOB) is a vacuum-compatible testbed at the University of Arizona built to advance and mature starlight suppression techniques in a space-like environment\cite{ashcraft_space_2022, van_gorkom_space_2022, ashcraft_space_2024, van_gorkom_space_2024, anche_space_2024}. In the configuration shown in Figure \ref{fig:scoob}, it consists of a charge-6 vector vortex coronagraph (VVC) equipped with a Boston Micromachines Kilo-C DM capable of achieving sub-10\textsuperscript{-8} contrasts at visible wavelengths. In this work, we report on our efforts in demonstrating a self-coherent camera (SCC) for HOWFSC on SCoOB. 

\section{THE SELF-COHERENT CAMERA}
\label{sec:scc}

\subsection{THEORY}
\label{sec:theory}

The SCC operates using coronagraph science images. By introducing an off-axis pinhole in the Lyot stop, a small amount of starlight is allowed to reach the science camera. This light interferes with the speckles and creates fringes. These fringes contain all the information needed to reconstruct the focal-plane electric field by isolating one of the sidebands in the optical transfer function (OTF) of the science image\cite{baudoz_self-coherent_2005, galicher_high-contrast_2013, galicher_self-coherent_2010, mazoyer_estimation_2013}. 

For a standard SCC, the center of the pinhole in the Lyot stop must be placed at a minimum separation of $1.5(D_{lyot} + D_{pinhole})$ from the center of the pupil. This is done so that the sidebands do not overlap with the central portion of the OTF. However, the post-Lyot stop optics on SCoOB are not oversized enough to support a pinhole at minimum separation. This causes the sidebands to overlap with the central portion of the OTF, which means that we need to modulate the pinhole. Using a 3D-printed flag mounted on a linear stage to open and close the pinhole, we can record a modulated science image with fringes or an unmodulated image without fringes. Similar to implicit electric field conjucation (iEFC), we can use these images to calculate a differential intensity image that acts as a proxy for the electric field to be used in HOWFSC\cite{haffert_implicit_2023}.

\subsection{MASK DESIGN}
\label{sec:design}

The Lyot pupil on SCoOB measured 9.1 mm in diameter. Because the post-Lyot plane optics were 12.7 mm in diameter, we could not place the SCC pinhole at the minimum separation needed to not overlap the sidebands with the central portion of the OTF. However, we still wanted to separate the pinhole as much as possible without vignetting the beam on the post-Lyot plane optics. Using a Fresnel model of SCoOB created using POPPY\cite{perrin_poppy_2016}, we simulated that we could offset the pinhole by 6 mm from the center of the Lyot pupil without vignetting on the post-Lyot plane optics. 

\begin{figure} [h!]
\centering\includegraphics[width=6cm]{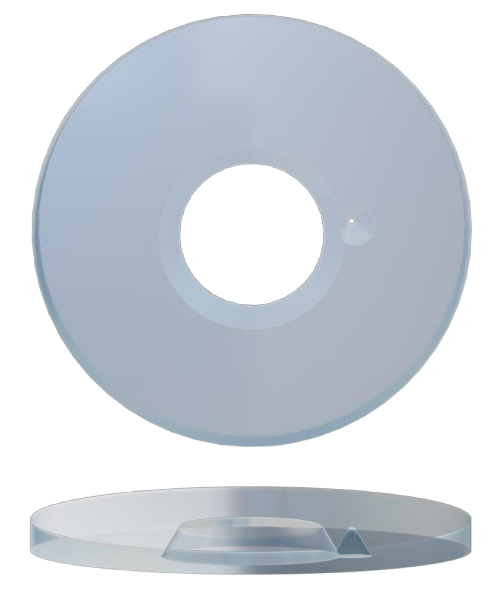}
\caption{CAD design of our SCC mask viewed from the back and side showing chamfered edges for the central stop and pinhole.} 
\label{fig:mask_design} 
\end{figure} 

We chose a 95\% Lyot stop and a 3\% SCC pinhole which measured 8.7 mm and 300 $\mu$m in diameter, respectively\cite{derby_tolerance_2022}. Using these numbers, we designed a 1.75-mm thick SCC mask. Both the central Lyot stop and the offset pinhole had a 30$^\circ$ off-normal chamfered edge to prevent light from scattering or coupling along the inside walls of the mask.

\subsection{MASK FABRICATION}
\label{sec:fab}

Our mask substrates were fabricated using a 1.75-mm thick wafer made of Schott Borofloat 33 glass. An ultrafast laser system was used to write a three-dimensional pattern throughout the thickness of the glass. To write the pattern, a STEP file was generated with the desired mask geometry and sliced using custom software to create a series of contours written at 10 $\mu$m depth increments. The ultrafast laser ($\lambda$ = 1030 nm) had a pulse energy of 2 $\mu$J, pulse duration of 7 ps full width at half maximum, circular polarization, and was focused using an objective lens with 0.25 numerical aperture. A constant repetition rate of 10 kHz was used for the pinhole while a rate of 25 kHz was used for all other features. The translation speed was set to 10 mm/s for the pinhole and 25 mm/s otherwise, resulting in laser pulses separated by 1 $\mu$m. 

\begin{figure} [h!]
\centering\includegraphics[width=5.9cm]{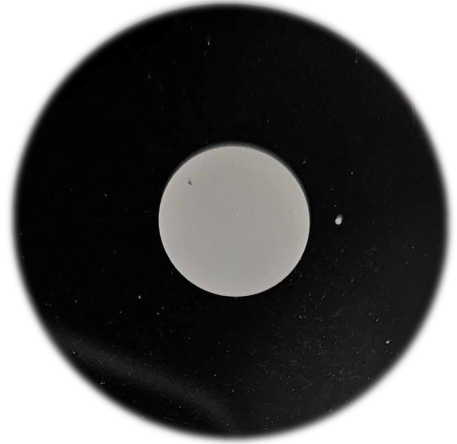}
\caption{Our final as-fabricated SCC mask viewed from the front coated with a layer of reflective aluminum.} 
\label{fig:mask_final} 
\end{figure} 

After writing, the substrate was then submerged in a potassium hydroxide (KOH) solution for roughly 4 hours until all features were etched. The KOH solution had an 8 mol/L concentration, was kept at 80$^\circ$C, and was stirred continuously during the entire etch. The final step after etching was to coat the front of the substrate with 100 nm of aluminum using a Temescal FC 2500 e-beam evaporator. Making the mask reflective allows us to perform reflective Lyot low-order wavefront sensing (LLOWFS) in tandem with HOWFS as shown in Milani et al. of these proceedings\cite{milani_llowfs_2025}. Future work will demonstrate concurrent operation of the SCC with LLOWFS.



\section{LABORATORY RESULTS}
\label{sec:results}

\subsection{CALIBRATION}
\label{sec:calib}

\begin{figure} [h!]
\centering\includegraphics[width=14cm]{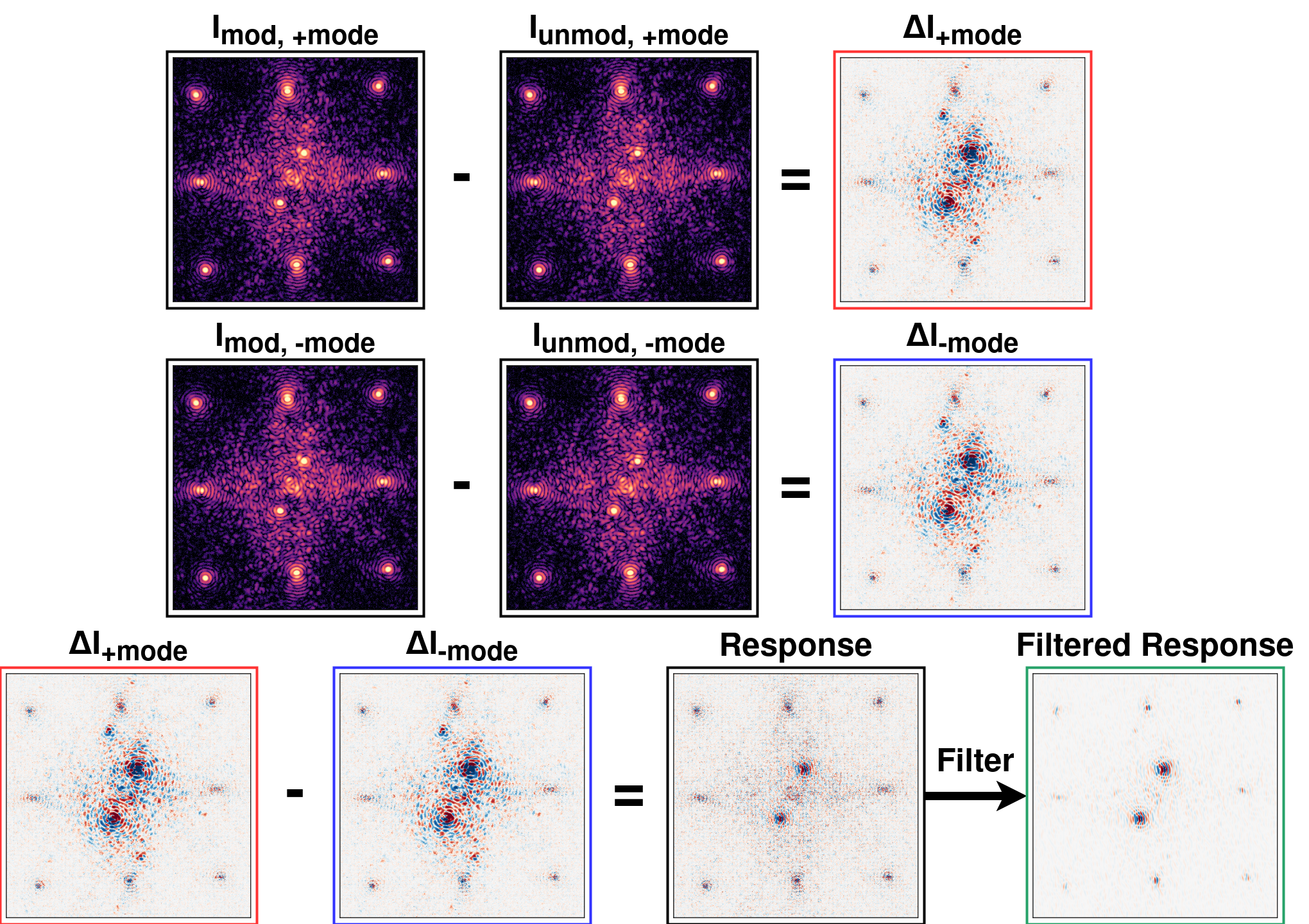}
\caption{Diagram of the Jacobian calibration procedure using our SCC. We used a two-sided Tukey window centered on the sidebands to filter the response.} 
\label{fig:calib} 
\end{figure}

Prior to digging the DH, we first needed to calibrate a jacobian which relates our DM control modes to their responses at the science camera focal plane. Any complete and independent basis can be used as DM control modes, but we chose to use Fourier modes since they concentrate the response within two $\lambda$/D-sized speckles which speeds up calibration. We used the calibration procedure described by Thompson et al\cite{thompson_performance_2022}. A schematic diagram of the procedure is shown in Figure \ref{fig:calib}. We applied each control mode as a positive and negative poke on the DM. For each poke, we captured a pair of modulated and unmodulated images. We used these to calculate a double differenced intensity response which was filtered using a two-sided Tukey window centered on the sidebands in the OTF to isolate the change in fringe structure. 

As shown in Figure \ref{fig:scoob}, we used a 3D-printed flag mounted on a linear stage to modulate the pinhole. However, we found that the stage we chose was not ideal because it was not encoded which led to repeatability issues when opening and closing the pinhole. To get around this, we mounted the flag such that it would close the pinhole when the stage was at one end of its travel range. To open the pinhole, we would translate the stage relative to the end of its range. To close the pinhole, we would purposefully send a command which would translate the stage past the end of its range. This ensured repeatability as we could verify the stage made it to the end of its travel range by listening to the motor as the stage hit the end of its range. However, this also made the process of taking a pair of modulated and unmodulated images very slow. 

During calibration, we needed to take a pair of modulated and unmodulated images for each positive and negative command poke. Because opening and closing the pinhole took several seconds with our linear stage, we opted to perform calibration by first opening the pinhole and taking only the modulated images for each positive and negative control mode poke. Then, we would close the pinhole and take the unmodulated images for each poke. This allowed us to calibrate jacobians roughly 15 times faster than if we had opened and closed the pinhole for each control mode. However, by taking our modulated and unmodulated images in two separate chunks, this left us vulnerable to testbed instabilities on minute-long time scales. Unfortunately, this summer we have noticed significant pointing drifts which would misalign the point spread function (PSF) on our charge-6 vector vortex focal plane mask during calibration. This caused a significant amount of calibration error which limited the achievable contrast when digging DHs using the SCC on SCoOB. While the source of these pointing drifts is currently unknown, in the future we hope that running LLOWFS during calibration will help eliminate these errors and allow us to achieve deeper contrasts with the SCC.

\subsection{MEASUREMENT}
\label{sec:measure}

\begin{figure} [h!]
\centering\includegraphics[width=14cm]{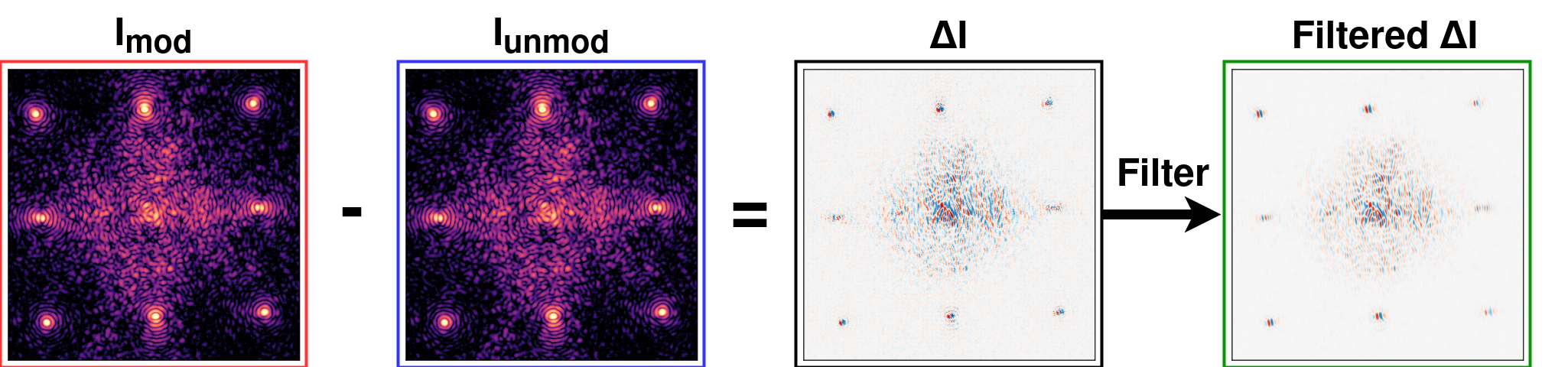}
\caption{Diagram of the measurement method using our SCC. We used a two-sided Tukey window centered on the sidebands to filter differential intensity measurements.} 
\label{fig:measure} 
\end{figure}

As mentioned previously, we used differential intensity measurements as a proxy for measuring the electric field using our SCC. A diagram showing the measurement procedure is shown in Figure \ref{fig:measure}. For each measurement, we took a pair of modulated and unmodulated images to obtain a differential intensity image. This was then filtered using a two-sided Tukey window centered on the sidebands in the OTF to isolate the change in fringe structure. We used the same window function for calibration and measurement. Using differential intensity is convenient as it leaves us robust to saturated pixels and other detector artifacts.

\subsection{DARK HOLE DIGGING}
\label{sec:digging}

\begin{figure} [h!]
\centering\includegraphics[width=11cm]{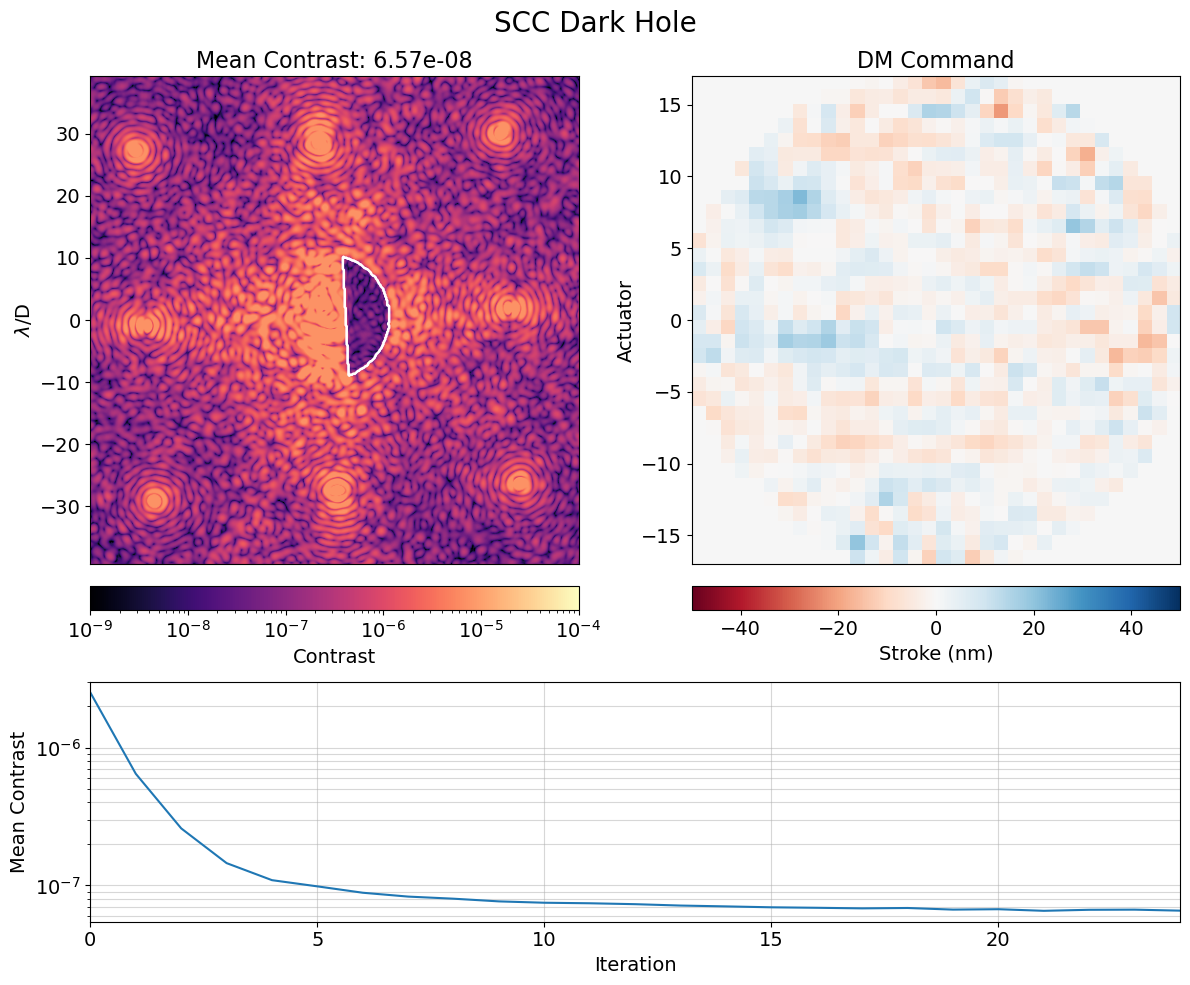}
\caption{(a) The dark hole dug using the SCC reached a final mean contrast of 6.57x10\textsuperscript{-8} with (b) the DM command shown. (c) The contrast curve shows the average contrast in the dark hole of the unmodulated image for each iteration of the algorithm.} 
\label{fig:dh_scc} 
\end{figure} 

\begin{figure} [h!]
\centering\includegraphics[width=11cm]{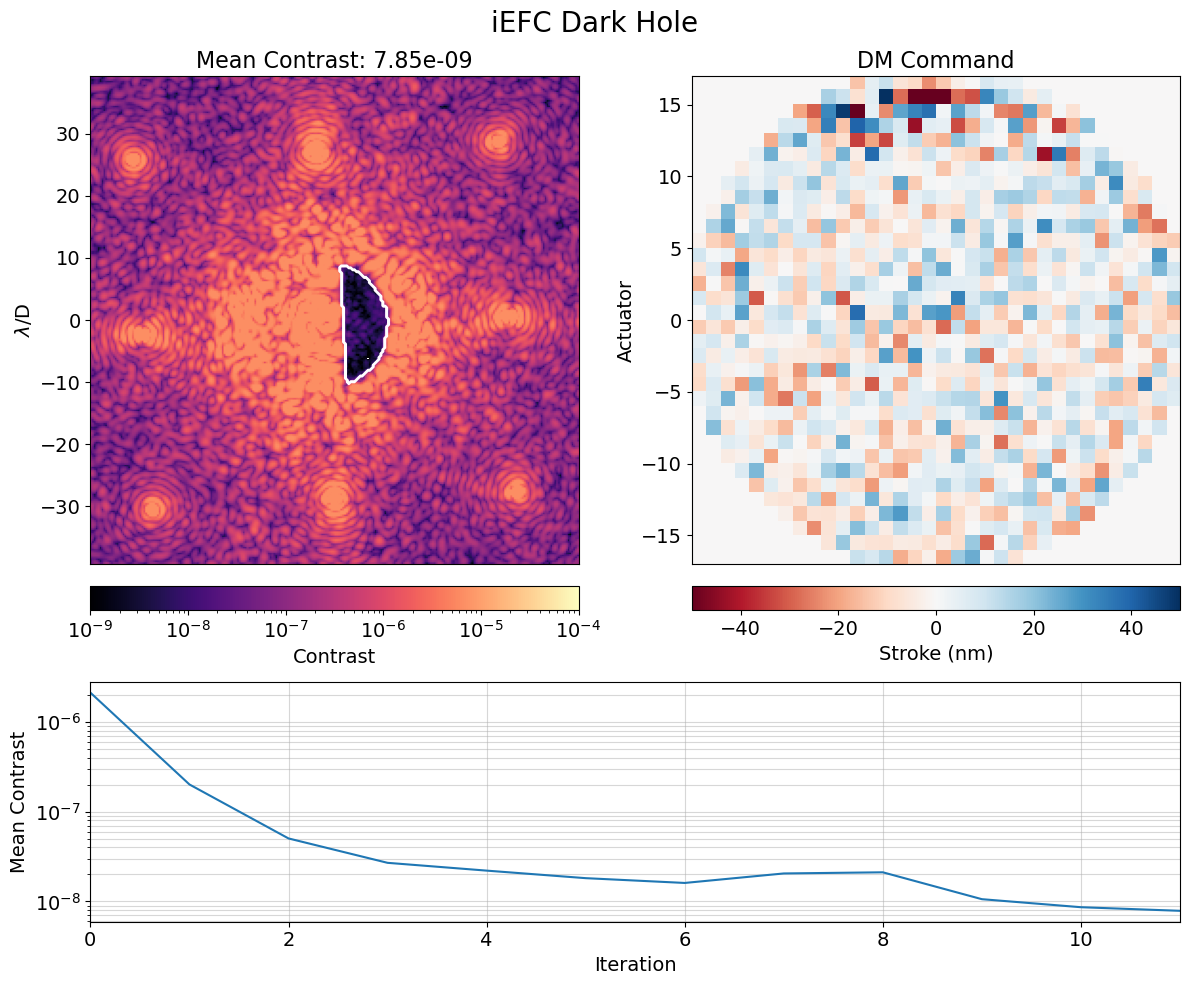}
\caption{(a) The dark hole dug using the iEFC reached a final mean contrast of 7.85x10\textsuperscript{-9} with (b) the DM command shown. (c) The contrast curve shows the average contrast in the dark hole for each iteration of the algorithm.} 
\label{fig:dh_iefc} 
\end{figure} 

Using the calibration and measurement procedures outlined above, we achieved a final mean contrast of 6.57x10\textsuperscript{-8} using the SCC in a 3 - 10 $\lambda/D$ half-annulus DH on SCoOB as shown in Figure \ref{fig:dh_scc}. DM commands were calculated by matrix multiplying differential intensity measurements with a control matrix created inverting the jacobian using Tikhonov regularization. While digging, we varied the regularization parameter in an attempt to let the control loop avoid local minima. However, we found that this rarely improved the final DM solution.

We check if DH contrast was limited by the SCC or the testbed, we compared the SCC DH against another DH dug using iEFC. Previous work has demonstrated that iEFC can reach sub-10\textsuperscript{-8} contrasts on SCoOB. Using similar calibration and measurement procedures, we found that iEFC was able to achieve a final mean contrast of 7.85x10\textsuperscript{-9} in a 3 - 10 $\lambda/D$ half-annulus DH. We believe these deeper contrasts are achievable using iEFC because it is less vulnerable to testbed drifts during calibration. Since iEFC uses DM probes to modulate the focal plane intensity, the modulated images used to calculate differential intensity can be obtained significantly faster than for the SCC. This let us take the images needed to calibrate a DM control mode sequentially instead of separately in two chunks spaced \~10 minutes apart.

\section{CONCLUSION}
\label{sec:conclusion}

In this work, we described the design and fabrication of an SCC mask for use on SCoOB. Using the SCC for HOWFS, we achieved a final mean contrast of 6.57x10\textsuperscript{-8} in a 3 - 10 $\lambda/D$ half-annulus DH. We baselined this result against iEFC, which achieved a final mean contrast of 7.85x10\textsuperscript{-9} in a 3 - 10 $\lambda/D$ half-annulus DH. We believe this difference is due to the SCC being more vulnerable to pointing drifts during calibration. Currently, we do not know the source of these new pointing drifts present on SCoOB. Future work will push for deeper contrasts with the SCC by diagnosing the cause of these drifts and correcting for them by running LLOWFS during SCC calibration. 

\acknowledgments 
 
Portions of this research were supported by funding from the Technology Research Initiative Fund of the Arizona Board of Regents and by generous anonymous philanthropic donations to the Steward Observatory of the College of Science at the University of Arizona. 

\bibliography{main} 

\begin{thebibliography}{10}

\bibitem{deeg_direct_2018}
Pueyo, L., ``Direct {Imaging} as a {Detection} {Technique} for {Exoplanets},'' in [{\em Handbook of {Exoplanets}}{\nolinebreak\hspace{0.1em}]},  Deeg, H.~J. and Belmonte, J.~A., eds.,  705--765, Springer International Publishing, Cham (2018).

\bibitem{guyon_extreme_nodate}
Guyon, O., ``Extreme {Adaptive} {Optics},''

\bibitem{giveon_electric_2007}
Give'on, A., Kern, B., Shaklan, S., Moody, D.~C., and Pueyo, L., ``Electric {Field} {Conjugation} - {A} {Broadband} {Wavefront} {Correction} {Algorithm} {For} {High}-contrast {Imaging} {Systems},''  {\bf 211},  135.20 (Dec. 2007).
\newblock ADS Bibcode: 2007AAS...21113520G.

\bibitem{ashcraft_space_2022}
Ashcraft, J.~N., Choi, H., Douglas, E.~S., Derby, K., Van~Gorkom, K., Kim, D., Anche, R., Carter, A., Durney, O., Haffert, S., Harrison, L., Kautz, M., Lumbres, J., Males, J.~R., Milani, K., Montoya, O.~M., and Smith, G.~A., ``The space coronagraph optical bench ({SCoOB}): 1. {Design} and assembly of a vacuum-compatible coronagraph testbed for spaceborne high-contrast imaging technology,'' in [{\em Space {Telescopes} and {Instrumentation} 2022: {Optical}, {Infrared}, and {Millimeter} {Wave}}{\nolinebreak\hspace{0.1em}]},  Coyle, L.~E., Perrin, M.~D., and Matsuura, S., eds.,  126, SPIE, Montréal, Canada (Aug. 2022).

\bibitem{van_gorkom_space_2022}
Van~Gorkom, K., Douglas, E.~S., Ashcraft, J.~N., Haffert, S., Kim, D., Choi, H., Anche, R.~M., Males, J.~R., Milani, K., Derby, K., Harrison, L., and Durney, O., ``The space coronagraph optical bench ({SCoOB}): 2. {Wavefront} sensing and control in a vacuum-compatible coronagraph testbed for spaceborne high-contrast imaging technology,'' in [{\em Space {Telescopes} and {Instrumentation} 2022: {Optical}, {Infrared}, and {Millimeter} {Wave}}{\nolinebreak\hspace{0.1em}]},  Coyle, L.~E., Perrin, M.~D., and Matsuura, S., eds.,  127, SPIE, Montréal, Canada (Aug. 2022).

\bibitem{ashcraft_space_2024}
Ashcraft, J.~N., Douglas, E.~S., Anche, R.~M., Van~Gorkom, K., Jenkins, E., Melby, W., and Millar-Blanchaer, M.~A., ``The space coronagraph optical bench ({SCoOB}): 3. {Mueller} matrix polarimetry of a coronagraphic exit pupil,'' in [{\em Space {Telescopes} and {Instrumentation} 2024: {Optical}, {Infrared}, and {Millimeter} {Wave}}{\nolinebreak\hspace{0.1em}]},  Coyle, L.~E., Perrin, M.~D., and Matsuura, S., eds.,  187, SPIE, Yokohama, Japan (Aug. 2024).

\bibitem{van_gorkom_space_2024}
Van~Gorkom, K., Douglas, E.~S., Milani, K., Ashcraft, J.~N., Anche, R.~M., Jenkins, E.~L., Ingraham, P., Haffert, S.~Y., Kim, D., Choi, H., and Durney, O., ``The space coronagraph optical bench ({SCoOB}): 4. vacuum performance of a high contrast imaging testbed,'' in [{\em Space {Telescopes} and {Instrumentation} 2024: {Optical}, {Infrared}, and {Millimeter} {Wave}}{\nolinebreak\hspace{0.1em}]},  Coyle, L.~E., Perrin, M.~D., and Matsuura, S., eds.,  74, SPIE, Yokohama, Japan (Aug. 2024).

\bibitem{anche_space_2024}
Anche, R.~M., Van~Gorkom, K., Ashcraft, J.~N., Douglas, E.~S., Haffert, S.~Y., and Millar-Blanchaer, M.~A., ``The space coronagraph optical bench ({SCoOB}): 5. {End}-to-end simulation of polarization aberrations,'' in [{\em Space {Telescopes} and {Instrumentation} 2024: {Optical}, {Infrared}, and {Millimeter} {Wave}}{\nolinebreak\hspace{0.1em}]},  Coyle, L.~E., Perrin, M.~D., and Matsuura, S., eds.,  189, SPIE, Yokohama, Japan (Aug. 2024).

\bibitem{baudoz_self-coherent_2005}
Baudoz, P., Boccaletti, A., Baudrand, J., and Rouan, D., ``The {Self}-{Coherent} {Camera}: a new tool for planet detection,'' {\em Proceedings of the International Astronomical Union}~{\bf 1},  553--558 (Oct. 2005).

\bibitem{galicher_high-contrast_2013}
Galicher, R., Mazoyer, J., Baudoz, P., and Rousset, G., ``High-contrast imaging with a self-coherent camera,''  88640M (Sept. 2013).

\bibitem{galicher_self-coherent_2010}
Galicher, R., Baudoz, P., Rousset, G., Totems, J., and Mas, M., ``Self-coherent camera as a focal plane wavefront sensor: simulations,'' {\em Astronomy and Astrophysics}~{\bf 509},  A31 (Jan. 2010).

\bibitem{mazoyer_estimation_2013}
Mazoyer, J., Baudoz, P., Galicher, R., Mas, M., and Rousset, G., ``Estimation and correction of wavefront aberrations using the self-coherent camera: laboratory results,'' {\em Astronomy \& Astrophysics}~{\bf 557},  A9 (Sept. 2013).

\bibitem{haffert_implicit_2023}
Haffert, S.~Y., Males, J.~R., Ahn, K., Van~Gorkom, K., Guyon, O., Close, L.~M., Long, J.~D., Hedglen, A.~D., Schatz, L., Kautz, M., Lumbres, J., Rodack, A., Knight, J.~M., and Miller, K., ``Implicit electric field conjugation: {Data}-driven focal plane control,'' {\em Astronomy \& Astrophysics}~{\bf 673},  A28 (May 2023).

\bibitem{perrin_poppy_2016}
Perrin, M., Long, J., Douglas, E., Sivaramakrishnan, A., Slocum, C., and {others}, ``{POPPY}: {Physical} {Optics} {Propagation} in {PYthon},'' {\em Astrophysics Source Code Library} ,  ascl:1602.018 (Feb. 2016).
\newblock ADS Bibcode: 2016ascl.soft02018P.

\bibitem{derby_tolerance_2022}
Derby, K., Haffert, S., Ashcraft, J.~N., Milani, K., Choi, H., Kim, Y.-S., Close, L., Mendillo, C., Chakrabarti, S., Allan, G., Pogorelyuk, L., Cahoy, K., N'Diaye, M., Kim, D., Males, J.~R., and Douglas, E.~S., ``Tolerance analysis of a self-coherent camera for wavefront sensing and dark hole maintenance,'' in [{\em Space {Telescopes} and {Instrumentation} 2022: {Optical}, {Infrared}, and {Millimeter} {Wave}}{\nolinebreak\hspace{0.1em}]},  Coyle, L.~E., Perrin, M.~D., and Matsuura, S., eds.,  153, SPIE, Montréal, Canada (Aug. 2022).

\bibitem{milani_llowfs_2025}
Milani, K., Van~Gorkom, K., Mendillo, C.~B., Anche, R., Ashcraft, J.~N., Derby, K., Males, J.~R., Schilperoort, A., and Douglas, E.~S., ``The space coronagraph optical bench (scoob): 6. demonstration of lyot low order wavefront control combined with high order wavefront control using a vortex coronagraph,'' in [{\em Techniques and Instrumentation for Detection of Exoplanets XII}{\nolinebreak\hspace{0.1em}]},  SPIE, San Diego, United States (Aug. 2025).

\bibitem{thompson_performance_2022}
Thompson, W.~R., Marois, C., Singh, G., Lardière, O., Gerard, B.~L., Fu, Q., and Heidrich, W., ``Performance of the {FAST} self coherent camera at the {NEW}-{EARTH} lab and a simplified {SCC} measurement algorithm,'' in [{\em Adaptive {Optics} {Systems} {VIII}}{\nolinebreak\hspace{0.1em}]},  Schmidt, D., Schreiber, L., and Vernet, E., eds.,  84, SPIE, Montréal, Canada (Aug. 2022).

\end{thebibliography}
\bibliographystyle{spiebib} 

\end{document}